\documentclass[rmp,nofootinbib,twocolumn,floatfix]{revtex4}
\usepackage{graphicx}
\ifnum\lefthyphenmin<2\lefthyphenmin=2\fi
\ifnum\righthyphenmin<2\righthyphenmin=2\fi
\bibpunct[, ]{[}{]}{,}{a}{,}{,}
\newcommand{\vS}{\vec{s}}
\newcommand{\sS}{{\sf s}}
\newcommand{\vsS}{\vec{{\sf s}}}
\newcommand{\cE}{{\cal E}}
\newcommand{\cS}{{\cal S}}

\newcommand{\nt}{\widetilde{n}}
\newcommand{\dos}{\Omega_\mathrm{sp}}
\newcommand{\Gns}{\Delta G_{\mathrm{ns}}}
\newcommand{\Eth}{E_{\mathrm{ns}}}
\newcommand{\betainv}{k_B T}
\newcommand{\kcal}{{\mathrm{kcal/mole}}}
\newcommand{\ecoli}{{\it E.~coli}}
\newcommand{\dG}{\Delta G}
\newcommand{\eps}{\varepsilon}
\newcommand{\bZ}{\overline{Z_\mathrm{b}}}
\newcommand{\Zsp}{\overline{Z_{\mathrm{sp}}}}
\newcommand{\bF}{\overline{F_\mathrm{b}}}
\newcommand{\Fb}{F_\mathrm{b}}
\newcommand{\Zb}{Z_\mathrm{b}}
\begin{document}
\title{Physical Constraints and Functional Characteristics of \\ Transcription 
Factor-DNA Interaction}
\author{Ulrich Gerland\footnote{These authors contributed equally to 
this work.}}
\author{J. David Moroz\footnote{Present address: NEC Research Institute,
4 Independence Way, Princeton, NJ 08540}\footnotemark[1]}
\author{Terence Hwa}
\affiliation{Department of Physics, University of California at San Diego, 
La Jolla, CA 92093-0319}
\date{\today}
%
%
\begin{abstract}
We study theoretical ``design principles'' for transcription factor-DNA 
interaction in bacteria, focusing particularly on the statistical
interaction of the transcription factors (TF's) with the genomic 
background (i.e., the genome without the target sites). 
We introduce and motivate the concept of {\em programmability}, i.e. 
the ability to set the threshold concentration for TF binding over 
a wide range merely by mutating the binding sequence of a target site. 
This functional demand, together with physical constraints arising from 
the thermodynamics and kinetics of TF-DNA interaction, leads us 
to a narrow range of ``optimal'' interaction parameters. 
We find that this parameter set agrees well with experimental data for the 
interaction parameters of a few exemplary prokaryotic TF's. 
This indicates that TF-DNA interaction is indeed programmable. 
We suggest further experiments to test whether this is a general feature 
for a large class of TF's. 
\end{abstract}
\maketitle
With rapid advances in the sequencing and annotation of entire genomes, 
the task of understanding the associated regulatory networks becomes 
increasingly prominent. 
Currently, many experimental and computational efforts are devoted to 
deciphering the genetic wiring diagram of a cell 
\cite{davidson02,eisen02,siggia01}.
Most of these efforts are focused on locating the functional DNA binding 
sites of transcription factors (TF's). 
This knowledge, together with the genomic sequences, will provide a 
qualitative picture of which gene products may directly affect the 
expression of which genes. 
While obtaining such wiring diagrams is tremendously important for the 
eventual understanding of gene regulation at the system level, 
this knowledge in itself is not sufficient for the quantitative
understanding of system-level effects. 
This has been dramatically shown in a detailed experimental study of the 
regulation of the {\it endo16} gene in Sea Urchin development 
\cite{davidson01}, which revealed an intricate regulatory function  
where a dozen or so TF's control the expression of a 
{\em single} gene. 
It would have been impossible to infer even the gross qualitative 
features of the transcriptional control from the knowledge of the 
binding sites alone. 

A major obstacle to progress is the lack of a {\em quantitative} 
understanding of the physical interaction between the TF's. 
However, even the simpler interaction between TF's and 
DNA sequences is not so well-understood quantitatively: It is common to
classify a potential TF-binding  DNA sequence in a 
``digital'' manner --- either the sequence is designated for TF-binding
or it is not. 
In this view of TF-DNA interaction, differences between the 
TF-binding sequences are only nuisances which impede straightforward 
bioinformatic methods of target sequence discovery. 
On the other hand, there are plenty of examples where differences between 
target sequences are known to be {\em functionally important} 
\cite{neidhardt}. 
In many cases, the binding of a TF to one site occurs 
only in the presence of some other TF, while the binding of the 
same TF to a different site does not require other TF's. 
This is often accomplished by differences in the binding sequences, and 
is believed to be the basis for combinatorial control and signal 
integration in gene regulation \cite{ptashne}. 
Also, different binding sites of the same TF can be ``tuned'' to bind 
at different TF concentrations, as suggested by a recent study of the 
\ecoli\ flagella assembly system \cite{alon}. 
If further experimental studies confirm that tuning of binding thresholds 
is indeed used genome wide to establish desired gene regulatory functions, 
then TF-DNA binding should be regarded more in an ``analog'' instead 
of a ``digital'' manner. 

In this work, we report our theoretical study on the ``design'' 
of TF-DNA interaction, assuming the analog scheme of 
operation. 
Specifically, we impose the functional requirement that the threshold 
concentration for TF binding to a site can be controlled over a wide 
range by the choice of the sequence alone; we refer to this as the 
``programmability'' of TF-DNA binding. 
Taken together with thermodynamic and kinetic constraints, this functional 
requirement leads to a narrow range of ``optimal'' TF-DNA interaction 
parameters. We then compare our result to experimentally known parameters 
for exemplary TF's to determine whether the design of these TF's 
would indeed {\em allow} the analog scheme of operation. 

To focus our discussion, we limit ourselves exclusively to 
the case of bacterial TF's which are the best
characterized experimentally. 
We study both the equilibrium occupancy of a target sequence, and the
dynamics of locating the target. 
Von Hippel, Berg, and Winter have already discussed many aspects of these 
issues in a series of seminal articles 
\cite{berg81,winter81a,winter81b,vonhippel86,berg87}.
Our study is  built firmly upon their work, but includes a number
of additional issues: (i) the effect of sequence-specific binding to 
the genomic background (non-target sequences) on the equilibrium 
occupation of a target sequence;
(ii) kinetic traps arising statistically from the genomic background,
and (iii) the desired programmability of TF-DNA binding. 
We adopt the model developed by von Hippel and Berg, and allow both
the sequence-specific and non-specific modes of TF-DNA binding.
Sequence-specific binding occurs if the binding sequence is 
sufficiently close to the best binding sequence, and is governed
quantitatively by a specificity parameter. 
For typical bacterial TF's whose binding sequences are 
no more than $15$ bases long, we find that our physical and functional 
requirements are best satisfied within a narrow regime of 
{\em intermediate specificity}, amounting to the loss of about $2 k_B T$ 
for each additional base mismatch from the best binding sequence. 
Furthermore, the kinetic constraint favors a low threshold to 
non-specfic binding, while the programmability requirement pushes the 
threshold to larger values. The optimal trade-off value depends only on 
the genome size, and lies about $16 k_B T$ above the energy of the best 
binding sequence for a genome of $10^7$ bases. 
These values correspond well with the interaction parameters of a number
of well-characterized TF's, which suggests that programmability
of TF-DNA binding is compatible with the reality of 
protein-DNA interaction and may be used by the organism 
to accomplish biological functions. 
We hope to stimulate further experiments determining the interaction 
parameters for a wider range of TF's (see `Discussion'). 
These experiments could either strengthen or falsify the programmibility 
concept, depending on whether the interaction parameters are generally 
in agreement with our prediction. 
\section*{Model of Transcription Factor-DNA Interaction}
Much of our knowledge on the details of TF-DNA 
interaction is derived from extensive biochemical experiments on a few 
exemplary systems, dating back to pioneering work in the 
late 70's \cite{dehaseth77,record77,kaohuang77,berg81,winter81a,winter81b} and 
continuing through recent years \cite{fields,stormo,sarai,takeda,frank97}. 
Furthermore, detailed structural information is available for many 
TF's from various structural families~\cite{pabo92}. 
Based on this knowledge, quantitative models of TF-DNA interaction 
have been established \cite{berg81,vonhippel86,berg87,stormo}. 
Together with the recent availability of genomic sequences, these models 
can be used to characterize the thermodynamics as well as the
dynamics of TF's with genomic DNA in a cell. 
We briefly review the primary model of TF-DNA interaction in this 
section, which serves to introduce our notation and formulate the problem. 

Biochemical and structural experiments, e.g. using {\it lac}-repressor 
\cite{record77,winter81a,frank97}, have established firmly 
that (i) TF's bind closely to the DNA
with a free energy $\Gns$ (with respect to the cytoplasm) regardless
of its sequence due to electrostatic interaction alone,
and (ii) additional sequence-specific binding energy 
can be gained (via hydrogen bonds) if the binding sequence 
is close to the recognition sequence of the TF.
Let the total binding (free) energy of a TF to a sequence 
$\vS = \{s_1, s_2, ..., s_L\}$ of $L$ nucleotides 
$s_i \in \{ {\tt A}, {\tt C}, {\tt G}, {\tt T}\}$ be $\dG[\vS]$ (with
respect to the cytoplasm), and let $\vS^*$ be the best binding sequence. 
$\dG[\vS]$ becomes sequence-independent, $\dG[\vS]\equiv\Gns$, 
if $\vS$ is far from $\vS^*$. 
This is believed to occur via a change in the conformation of the
TF, from one which allows more hydrogen bond formation 
to another which brings the positive charges of the TF closer to 
the negatively charged DNA backbone \cite{winter81b}.

For this study, it will be convenient to measure all energies with
respect to that of the best binder, $\dG[\vS^*]$. Let us define
$E[\vS] \equiv \dG[\vS]-\dG[\vS^*]$.
Furthermore, we will introduce the threshold energy 
$\Eth \equiv \Gns - \dG[\vS^*]$
where TF-DNA binding switches from the specific to the non-specific
mode (for {\it lac}-repressor, $\Eth \approx 10\,\kcal$).
Then given the above model of TF-DNA interaction, 
and assuming that the TF is bound to the DNA 
essentially all the time\footnote{{\it In vivo} measurements for the 
case of {\it lac}-repressor found less than 10\% of the TF's 
were unbound \cite{kaohuang77}. This agrees well with an estimate based 
on a typical prokaryotic cell volume of 3~$\mu{\rm m}^3$, a genome length 
of $5\cdot 10^6$ bases, and a non-specific binding constant on the order 
of $10^4\,{\rm M}^{-1}$ under physiological conditions \cite{dehaseth77}, 
which yields a fraction of unbound TF's at a few percent level.}, 
all thermodynamic quantities regarding
this TF can be computed from the partition 
function\footnote{One should also include the reverse-complement
of the genomic sequence in the evaluation of the 
partition function $Z$. In order not to make the notation too 
complicated, we extend the definition of ``genomic sequence'' to 
include its complement.}
\begin{equation}
  Z = \sum_{j=1}^N e^{- \beta E[\vsS_{j}]} \, + \, N\cdot e^{-\beta \Eth},
  \label{Z}
\end{equation}
where $\beta^{-1} = k_BT \approx 0.6\,\kcal$ and
$\vsS_j$ denotes the subsequence of the genomic sequence $\{\sS_1, \sS_2,...,
\sS_N\}$ from position $j$ to $j+L-1$. The binding length of a typical
bacterial TF is $L= 10 \sim 20$ basepairs (bp).
The length of genomic sequence $N$ is typically several million bp.

The form of the binding energy $E[\vS]$ has been studied experimentally 
for several TF's \cite{fields,stormo,sarai,takeda}.
In particular, recent experiments on the TF Mnt from 
bacteriophage P22~\cite{fields} support the earlier model \cite{vonhippel86} 
that the contribution of each nucleotide in the binding sequence to the 
total binding energy is approximately independent and additive, i.e., 
\begin{equation}
  \label{energy}
  E[\vS] = \sum_{i=1}^L \cE_i(s_i)\;.
\end{equation}
For the TF's Mnt, Cro, and $\lambda$-repressor, the parameters of the 
``energy matrix'' $\cE_i(s_i)$ have actually been determined experimentally 
by {\it in vitro} measurements of the equilibrium binding constants 
$K[\vS]\propto e^{-\beta E[\vS]}$ for every {\em single-nucleotide}
mutant of the best binding sequence $\vS^*$~\cite{fields,sarai,takeda}. 
Due to our definition of the energy scale, $\cE_i(s_i)=0$ for
$s_i=s_i^*$ and $\cE_i(s_i)>0$ for $s_i\neq s_i^*$; the latter will
be referred to as  ``mismatch energies''.
While the simple form of the binding energy (\ref{energy}) will certainly 
not hold for all TF's, and di-, tri-nucleotide 
correlation effects are likely to be important in many cases 
(e.g., to some extent for {\it lac}-repressor \cite{frank97}), 
the key results of our study are not sensitive to such correlations
as long as there is a wide range of binding energies for different
binding sequences. Thus we will adopt the simple form (\ref{energy}) 
for this study.
For the three well-studied TF's, 
the mismatch energies are typically in the range of $1\sim 3\, k_BT$'s.
While the threshold energies $\Eth$ have not been carefully measured
for these TF's, it is believed that non-specific binding does not
occur until the binding sequences are at least $4$ to $5$ mismatches
away from $\vS^*$ (G.~Stormo, private communication). 
\section*{Genomic Background and Target Recognition}
\subsection*{Thermodynamics} 
Let us first consider the binding of a {\em single} TF
to its target sequence, denoted by $\vS_t$. We will assume that thermal
equilibrium can be reached within the relevant cellular time scale
and discuss the important kinetics issue afterwards.
The effectiveness of the binding of the TF to its target is then
described by the equilibrium binding probability $P_t$, which depends not 
only on the binding energy $E_t \equiv E[\vS_t]$, but also on the 
interaction with the rest of the genomic sequence. 
Let the contribution of this genomic background to the partition function
be $\Zb$, then the binding probability to the target is given by 
\begin{equation}
  \label{bind_prob}
  P_t=\frac{1}{1+e^{\beta (E_t-\Fb)}}\;,
\end{equation}
where $\Fb = -\betainv \ln \Zb$ is the effective binding energy 
(or free energy) of the entire genomic background. 
Eq.~(\ref{bind_prob}) is a sigmoidal function of $E_t$ with a (soft) 
threshold at $\Fb$, i.e., a TF binds (with probability $P_t>0.5$) 
if $E_t<\Fb$. Since $E_t \ge 0$ by definition, we must have 
\begin{equation}
\Fb \ge 0
\label{cond.1TF.a}
\end{equation}
in order for a target sequence to be recognized by a single TF 
(we consider multiple TF's below).

\begin{figure}
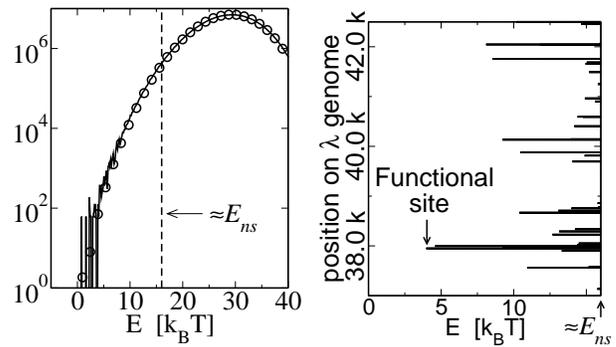

\includegraphics[width=4cm]{fig1a}
\includegraphics[width=4cm]{fig1b}
\caption{\protect\flushing \label{rem_applies}
For the purpose of TF binding, the genome may 
be treated as random DNA plus functional target site(s): 
(a) Histogram of the specific binding energies for Cro [solid line] 
on the \ecoli\ genome, together with the average histogram [circles] for 
Cro on random nucleotide sequences (synthesized with the same length 
and single-nucleotide frequencies as the \ecoli\ genome; normalization for 
both histograms such that maximum is at $N$). 
Except for statistical fluctuations at the low energy end, the histograms 
are indistinguishable from each other. The approximate position of the
threshold energy for nonspecific binding $\Eth$ is indicated 
as the thin straight line.
(b) Energy landscape for Cro on the bacteriophage $\lambda$ DNA. 
The landscape appears to be {\em random}, e.g. no ``funnel'' guides the 
TF to the target site.
The spatial correlation function of the landscape (not shown) decays
quickly to zero beyond the scale of $L=17$ for this case. 
Random energy landscapes are also found for the other two TF's with 
known energy matrices (not shown). 
}
\end{figure}

The background contribution can be computed for
any given TF and genome according to Eq.~(\ref{Z})
if the binding energy matrix, the threshold energy $\Eth$, 
and the genomic sequence is known. We will instead seek a description
that is independent of the specifics of the genomic sequences and
energy matrices. To accomplish this, we observe first that
for the few well-studied TF's, the interaction of
the TF with the genomic background  can be well approximated by the
interaction of the TF with {\em random} nucleotide sequences
of the same length and single-nucleotide frequencies $p(s)$.
This is illustrated in Fig.~\ref{rem_applies}(a) where the histogram 
of binding energies
obtained by using the binding energy matrix $\cE_i(s)$ for the 
TF Cro on the \ecoli\ genome (solid line) coincides well with
the histogram of the same energy matrix applied to random nucleotide
sequences (circles). Moreover, there appears to be hardly any positional
correlation in the binding energies along the genome, as shown by the
``energy landscape'' in Fig.~\ref{rem_applies}(b) 
(see caption for details). In the following, we will therefore
describe the effect of the genomic background by treating it as a 
random nucleotide sequence for a generic TF. In particular,
we will describe the genomic background partition function 
by $\Zb = Z_\mathrm{sp} + N\cdot e^{-\beta\Eth}$
where the contribution due to sequence-specific binding is
\begin{equation}
  \label{Zsp.def}
  Z_\mathrm{sp} = \sum_{\vS \in \cS(N)} e^{-\beta E[\vS]},
\end{equation}
with $\cS(N)$ denoting a given collection of $N$ random nucleotide
sequences of length $L$, drawn according to the frequency 
$p(s)$ for each nucleotide $s$. 

Even with the random sequence approximation (\ref{Zsp.def}), computation of the
background energy $\Fb = -\betainv \ln \Zb$ is nontrivial in principle:
From its definition, it is clear that $\Fb$ is a random variable,
and its precise value will depend on the actual collection of 
sequences $\cS(N)$.
We are interested in the {\em typical} value of $\Fb$, a reasonable 
approximation of which is its statistical average,
$\bF \equiv -\betainv\,\overline{\ln \Zb}$.  
[We use an overbar to denote averages over an ensemble of different 
sequence collections $\cS(N)$.]
Computing the average $\overline{\ln \Zb}$ is however difficult to do 
for an arbitrary energy matrix $\cE_i(s)$ short of performing numerical 
simulations. 
An alternative is to compute the ensemble average of $\Zb$, 
i.e., $\bZ = \Zsp + N\, e^{-\beta\Eth}$, where
\begin{equation}
  \label{Zsp}
  \Zsp= N \, \prod_{i=1}^{L}\left[\sum_{s=\{{\tt A,C,G,T}\}}
   e^{-\beta\cE_i(s)} \; p(s)\right] 
\end{equation}
with the single-nucleotide frequencies $p(s)$, and assume  that
\begin{equation}
  \label{anneal}
  \Fb \approx - \betainv\, \ln \bZ. 
\end{equation} 
This is, for example, the approach taken by Stormo and 
Fields~\cite{stormo} in their analysis of the TF 
Mnt\footnote{In Ref.~\cite{stormo}, the nonspecific binding was not
included so that $\bZ = \Zsp$, 
and the energy scale was shifted such that $\bZ = N$.}.
We note in passing that $\bZ$ can be more compactly written in terms
of the density of states $\dos(E)$ for specific binding [the normalized
version of the histogram in Fig.~\ref{rem_applies}(a)], i.e., 
\begin{equation}
\bZ =  N \cdot \sum_E\, \dos(E)\, e^{-\beta E} \; 
+ \; N \cdot e^{-\beta \Eth}.
\label{bZ}
\end{equation}

The relation (\ref{anneal}) is based on the so-called ``annealed
approximation'' $\overline{\ln Z} \approx \ln \overline{Z}$ which is valid  
for the genomic sequence length $N \to \infty$ 
but not always appropriate for finite $N$, e.g., if the partition function 
is dominated by a few low energy terms.
Much is known from statistical physics about systems of the type defined 
by the partition function $Z_\mathrm{sp}$ in (\ref{Zsp.def}), 
generically known as the 
Random Energy Model or REM\footnote{In many applications, 
including protein folding~\cite{bryngelson}, the REM was introduced
to {\em approximate} the random background interaction. 
The TF-DNA interaction as defined by (\ref{Zsp.def}) 
represents one of the few systems for which the REM description is 
{\em directly applicable}.}, introduced by Derrida \cite{derrida81}.
It turns out that the annealed approximation is valid as long as 
the system's entropy $S$ is significantly larger than zero, reflecting 
the contribution of many terms in the partition sum. 
We will see further below that proper function of the TF's requires the 
system to be in a regime where the annealed approximation is safely 
applicable. We will thus take the relation (\ref{anneal}) for granted. 
In this case, the condition (\ref{cond.1TF.a}) for the recognition 
of the target sequence by a single TF becomes
\begin{equation}
 \bZ \le 1.
 \label{cond.1TF.b}
\end{equation}
\subsection*{Search Dynamics}
In order to carry out their function properly, TF's not 
only need to have a high equilibrium binding probability to their 
targets, but also must be able to locate them in a reasonably 
short time (e.g. less than a few minutes) after they have been 
activated by an inducer or freshly produced by a ribosome. 
This constitutes a constraint on the ``search dynamics'' of 
TF's. 

In their non-specific binding mode, TF's are still  
strongly associated with the DNA, but are able to diffuse (i.e., slide) 
randomly along the genome \cite{berg81,winter81a,winter81b}. 
However, pure one-dimensional diffusion would be an inefficient 
search process, since it is very redundant (e.g., a 1D random walker 
always returns back to the start.)
For instance, assuming generously a 1D diffusion constant of 
$D_1 \approx 1\,\mu{\rm m}^2/{\rm sec}$ \cite{winter81b},
one finds a time $T_\mathrm{1D} \sim N^2/D_1 \sim 10^6\, {\rm sec}$ 
for a single TF to diffuse around a bacterial genome of length 
$N\approx 5\times 10^6$ bp (about $1\, {\rm mm}$). 
Thus, in order to find a target within a few minutes via 1D diffusion, 
one would need at least 100 TF's per cell to search in parallel 
(so that the search length $N$ is reduced by a factor of $100$). 
On the other hand, there are well-documented examples where regulation 
is accomplished effectively by only a few TF's in a cell 
(e.g., about 10 for {\it lac}-repressor in {\it E. coli}) \cite{droge}.

\begin{figure}
\includegraphics[width=4.0cm]{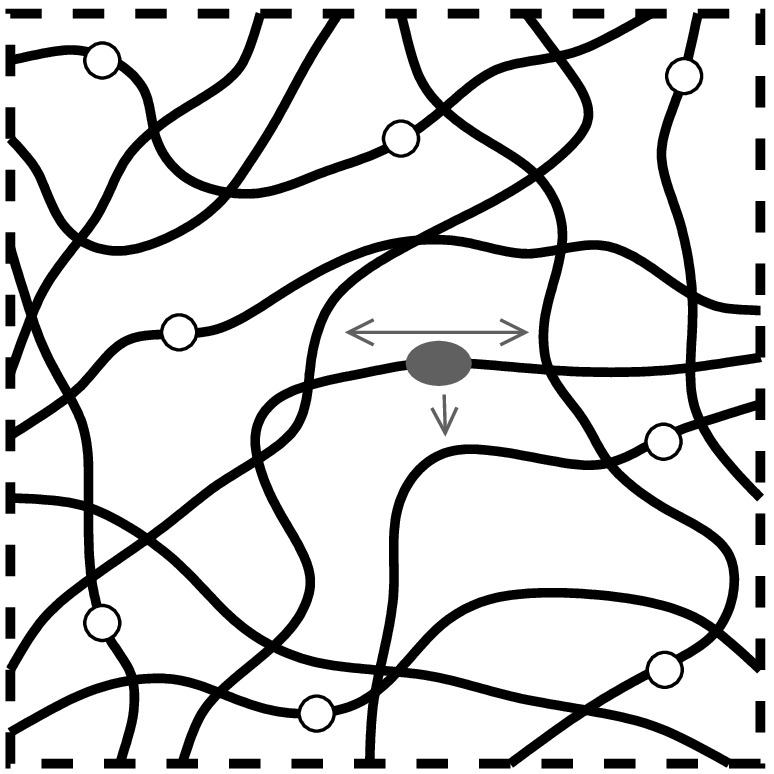}
\hspace*{0.2cm}
\raisebox{-0.1cm}{\includegraphics[width=3.8cm]{fig2b}}
\caption{\protect\flushing \label{fig2}
(a) Schematic illustration of the search dynamics: a TF 
(represented by a solid ellipse) 
moves among genomic DNA (lines) via a combination of
1D diffusion (along the genome) and 3D diffusion (hopping 
between nearby segments) as illustrated by the arrows.
The open circles indicate the potential kinetic traps
which are sites that are preferred by the TF
in a random background.
(b) Dependence of the chemical potential $\mu$ on the number $n$ of 
TF's in a cell for Mnt, Cro, and $\lambda$-repressor,
obtained by directly solving and inverting the defining equation
(\ref{mu}). The comparison with the dashed line $\mu=\betainv \ln n$ 
shows that $\mu(n)$ is sufficiently well described by the simple 
expression (\ref{mu.1}) over the regime $1< n < 1000$. 
}
\end{figure}

As studied in detail by Winter, Berg, and von Hippel 
\cite{berg81,winter81a,winter81b}, the search dynamics of TF's involves 
instead a combination of sliding along the DNA at short 
length scales and {\em hopping} between different segments of DNA 
(either over the dissociation barrier through the cytoplasm, or by direct 
intersegment transfer); see Fig.~\ref{fig2}(a). 
This search mode is much faster (given the high DNA concentration inside 
the cell), since the dynamics is essentially 3D diffusion beyond the 
hopping scale, and 3D diffusion is much less redundant than 1D diffusion. 
For example, if the TF's were not bound to the DNA at all, 
a single TF of a few nm in linear dimension $\ell$ would locate its 
target in a cell volume $V_{\rm cell}$ of several $\mu{\mathrm m}^3$ in 
the average first passage time of 
$T_\mathrm{3D}=V_{\rm cell}/(4\pi\ell D_3) \sim 10$ sec, 
given a 3D diffusion constant on the order of 
$D_3\sim 10\,\mu{\mathrm m}^2/{\rm sec}$ \cite{elowitz99}. 
The search time $T_\mathrm{3D/1D}$ for the combined 1D/3D diffusion 
under {\it in vivo} conditions can be estimated to be comparable to 
$T_\mathrm{3D}$ \cite{winter81b}. 
Hence, the search time is short enough to comfortably allow even a single 
TF to locate its target within the physiological time scale. 

In the study of the search dynamics reviewed above, binding of 
the TF to the genomic background was assumed to occur at a single 
energy value, namely, the non-specific energy $\Gns$~\cite{berg81}. 
On the other hand, the ``energy landscape'' of Fig.~\ref{rem_applies}(b) 
clearly shows that the random genomic background contains many isolated 
sites with binding energies far below $\Gns$. 
These sites constitute kinetic traps which can, in principle, drastically 
impede the local search process, if the energy difference to their 
surroundings is sufficiently large\footnote{Note that the additional 
sequence-specific binding energy to a `spurious site' in the background 
equally increases the kinetic barrier for sliding to a neighboring site 
as well as for dissociation into the cytoplasm.}. 
Thus to fully understand the search dynamics, we need to characterize 
the effect of kinetic traps in the genomic background: what is the 
constraint on the design of TF-DNA interaction imposed by 
requiring that the effect of kinetic traps be negligible? 

At each binding sequence $\vS_j$ with energy $E_j \equiv E[\vS_j]<\Eth$, 
the TF typically spends a time 
$\tau_j = \tau_0 \cdot e^{\beta(\Eth-E[\vS_j])}$, where $\tau_0$ is the 
average `waiting time' of the TF at a non-specific binding site. 
Along the search path of the TF, the {\em average waiting time} 
$\overline{\tau}$ per binding site is then simply given by
\begin{equation}
 \overline{\tau}=  \tau_0 
 \sum_E \,\dos(E)\,\big[ e^{\beta(\Eth-E)} \theta(\Eth-E) + \theta(E-\Eth) 
 \big].
 \label{tau-ave}
\end{equation}
Here we assumed as before that the genomic sequence is random so that 
the sequence-specific binding energy $E$ 
can be treated as a random
variable drawn from the distribution $\dos(E)$. The second term,
with the help of the unit step function
$\theta(x)$, is used to express the fact that there is no kinetic
trap for the (majority of) sites with $E>\Eth$.

A comparison of (\ref{tau-ave}) and Eq.~(\ref{bZ}) for the average
partition function $\bZ$ immediately yields the important 
relation\footnote{Note that this relation is actually independent
of the additive form of the binding energy (\ref{energy}).}
\begin{equation} 
\overline{\tau}/\tau_0 \approx \bZ\, e^{\beta\Eth}/N,
\label{trap}
\end{equation}
since in Eq.~(\ref{bZ}), the second term dominates for $E> \Eth$.
As expected, the kinetic trap factor $\overline{\tau}/\tau_0$ 
grows exponentially with $\Eth$, the threshold to nonspecific binding.
On the other hand, we note from $\bZ = \Zsp + N\, e^{-\beta\Eth}$ 
(see `Thermodynamics') that the trap factor 
can be made to be of order one so that the dynamical analysis of 
Refs.~\cite{berg81,winter81a,winter81b} remains qualitatively valid, if 
$\Zsp \le N\, e^{-\beta\Eth}$. 
The physical meaning of this condition is that the average effect
of the kinetic traps can be rendered small if the sum of the waiting 
times does not exceed the order of the plain diffusion time.
As we will see, this can be accomplished by choosing the binding energy 
matrix $\cE_i(s)$ and $\Eth$ appropriately. 
Combining this kinetic constraint with Eq.~(\ref{cond.1TF.b}), 
we obtain the condition
\begin{equation}
  \Zsp \le N\, e^{-\beta\Eth} \le 1
  \label{cond.1TF}
\end{equation}
for the rapid recognition of a target sequence by a single TF.
\section*{Programmability of Binding Threshold}
\subsection*{Multiple transcription factors}
There are of course typically multiple copies of the same TF in the cell, 
and the regulatory function is accomplished
if any one of these TF's binds to the target sequence. 
If the cell contains $n$ copies of a given TF, then the
occupation probability for the target sequence, Eq.~(\ref{bind_prob}), 
is replaced by the Fermi distribution (or `Arrhenius function') 
$P_t=1/(1+e^{\beta (E_t-\mu)})$,
since each binding sequence can at most be occupied by one TF. 
The chemical potential $\mu(n)$ is determined implicitly from 
the condition\footnote{Here, the exclusion between overlapping binding 
sites can be neglected since $n\ll N$. Also, we have not included the 
(unimportant) exclusion between the specific and the unspecific binding 
mode at a given site.} 
\begin{equation}
  n = \sum_E \; \left[\dos(E)+\delta_{E,\Eth}\right] 
  \cdot \frac{1}{1+e^{\beta (E-\mu)}},
\label{mu}
\end{equation}
where the quantity in brackets represents the total density of states.
In the simplest scenario, where steric exclusion between TF's bound 
to the non-target sequences is negligible, one has \cite{vonhippel86}
\begin{equation}
  \mu \approx \betainv \ln n + \Fb\;.
  \label{mu.1}
\end{equation}
This is empirically found to be a good approximation for those 
TF's with known binding energy matrices, as shown in 
Fig.~\ref{fig2}(b).
We will adopt the form (\ref{mu.1}) for the chemical potential
of a generic TF in this study; a general argument will 
be given later to justify this choice, even for the case where mulitple 
target sequences are present in the same genome.

Using (\ref{mu.1}), the occupation probability  
can be more succinctly written as $P_t = 1/(1+\nt_t/n)$, where
\begin{equation}
  \nt_t = e^{\beta(E_t - \Fb)} 
  \label{nt}
\end{equation}
denotes the (soft) {\em threshold concentration} of the TF for occupation 
of the target sequence. 
\subsection*{Programmability}
The allowed values of the background free energy $\Fb$ for the
binding of the target sequence obviously depend on the TF 
concentration $n$. For example, we have the condition (\ref{cond.1TF.a})
for $n=1$, while smaller values are allowed for $n>1$.
It thus appears that the allowed $\Fb$'s are different for the
different TF's, since they would typically be present in the cell 
with different concentrations. On the other hand, even for a given
TF species, the {\em desired} binding threshold  
may not be at a {\em single} concentration for different target sites, 
but can {\em vary} depending on functional demands.  
For example, it can be desirable 
to turn on different genes/operons at different TF concentrations,
in order to maintain a {\em temporal order} in the expression
of different operons as the concentration of the controlling TF 
gradually changes over time. This effect was observed recently for the
\ecoli\ flagella assembly~\cite{alon} and SOS response systems
(U.~Alon, private communication).

As another example, 
consider the case where a particular TF A
is involved in the regulation of two operons, X and Y. Suppose 
it is desired that A activates the transcription of operon X on
its own at a concentration $n_A$, while operon Y should {\em only}
be activated if A is present (at the same concentration $n_A$) 
{\em together} with another TF B that can bind cooperatively with A. 
It is then desirable to have a strong binding site for A in the regulatory
region of operon X such that its threshold $\nt_{A,X} < n_A$, 
and a weak binding site in the regulatory 
region of operon Y, with a threshold $\nt_{A,Y} > n_A$.
The latter insures that the operon Y will not be accidentally
activated by fluctuations in $n_A$ alone, and only when
the TF B is present would  the attractive interaction between
A and B induce the two to bind to their targets.

The above examples show that it is functionally desirable 
to have the ability to {\em set} the binding threshold $\nt_t$ of a
given TF to each of its target sequence $\vS_t$ 
{\em individually}. As is clear from the defining expression (\ref{nt}), 
this can only be done through the choice of the target sequence 
$\vS_t$ which affects $E_t$, since the other variable, $\Fb$, is fixed 
for a given TF. 
We refer to the ability to control the binding threshold $\nt_t$
through the choice of the target sequence $\vS_t$ {\em alone} as 
``programmability'' of the binding threshold. Assuming that programmability
is a desirable feature of TF-DNA interaction
(since sequence changes can be easily accomplished by point mutation
if the functional need arises), we seek
to determine the specifics of the TF-DNA interaction, e.g., the 
binding matrix $\cE_i(s)$, the length of the binding sequence $L$, 
and the threshold energy $\Eth$, which allow the targets to be 
{\em maximally programmable}. 
\subsection*{Two-state model and parameter selection}
Specifically, let us require programmability of the binding threshold 
over the entire range $\nt = 1 \ldots 10^3$, since typical cellular 
TF concentrations range from a few to a few hundred per cell. 
The lower bound $\nt\approx 1$ immediately imposes the condition 
(\ref{cond.1TF.a}) on $\Fb$, or, taking also the kinetic constraint 
into account, the condition (\ref{cond.1TF}). 
Furthermore, in order to tune $\nt$ throughout the desired range with a 
reasonable resolution, it is necessary to have the ability to change 
$E_t$ from $0$ to $\betainv \ln 10^3 \approx 7 \betainv$,
in small increments. This requires the non-zero entries of the binding
energy matrix $\cE_i(s)$ to take on small values. 
Which choices for the TF-DNA interaction parameters 
($\cE_i(s)$, $L$, $\Eth$) can simultaneously satisfy the latter 
requirement and condition (\ref{cond.1TF})? 

The combined effect of these physical constraints and functional demands 
is best understood by simplifying the energy matrix $\cE$ such that 
we retain the essential and generic aspect of sequence-specific binding, 
while eliminating all TF-specific details. Towards this end,
we adopt the two-state model originally introduced by von Hippel and 
Berg \cite{vonhippel86}, characterizing all of the non-zero entries of the 
{\em significant positions}\footnote{Note that the energy matrices for most
TF's contain a number of (fixed) positions which have no strong preference
for any of the nucleotides. We will not consider these positions
in the ensuing discussion of the two-state model, and  will use $L$
to refer to the total number of significant positions.} in the energy
matrix by a {\em single} value, i.e., 
\begin{equation} 
  \label{twostate}
  \beta\cE_i(s) ~ = ~ \left\{
  \begin{array}{ll}
    0 & {\rm if}~s=s_i^*\\
    \eps>0 & {\rm if}~s\neq s_i^*
  \end{array}\right. \;,
\end{equation}
where $\eps$ is a dimensionless ``discrimination energy'' 
(in units of $k_BT$).
It describes the energetic {\em preference} of the TF
for the optimal binding sequence $\vS^*$, and is a crucial parameter
controlling the specificity of the TF.
Within the two-state model, the binding energy to the target $\vS_t$ is 
simply $\eps$ times the total number of mismatches between the target
and the best binder $\vS^*$, i.e., 
$E[\vS_t] = \eps \cdot |\vS_t - \vS^*|$, where $|...|$ denotes the
Hamming distance between two sequences. Clearly, programmability 
is best satisfied with a small $\eps$ which enhances the resolution
of the programmable binding threshold.

\begin{figure}
\includegraphics[width=3.75cm]{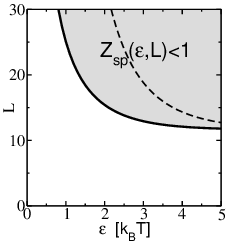}
\hspace*{0.2cm}
\raisebox{0.1cm}{\includegraphics[width=3.75cm]{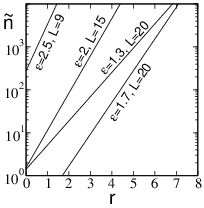}}
\caption{\protect\flushing \label{cond.Zsp}
(a) Plot of the region where $\Zsp(\eps, L) \le 1$. The boundary 
$L^*(\eps)$ for $N=10^7$ is indicated by the solid line; see text.
The dashed line $\ln(N)/(\ln \zeta^{-1}(\eps)-\eps/(1+e^\eps/3))$ 
indicates the onset of the glass transition in the 
random energy model where the annealed approximation breaks down. 
As argued in the text, the desired parameter regime is close to 
$\Zsp =1$, so that the annealed approximation is justified. 
(b) The binding threshold $\nt$ as a function of the total number of 
mismatches $r$ of the target sequence $\vS_t$ from the best binder $\vS^*$
at different parameter combinations $(\eps,L)$.
}
\end{figure}

The two-state model (\ref{twostate}) also allows an explicit evaluation 
of the condition (\ref{cond.1TF}) via the formulae (\ref{Zsp})
for $\Zsp$. Assuming for simplicity equal single-nucleotide
frequencies in the background (i.e., $p(s)=1/4$), 
the quantity in the bracket of (\ref{Zsp}) is easily evaluated.
We have $\Zsp(\eps,L) = N\cdot \zeta^L(\eps)$ where
$\zeta \equiv \sum_s e^{-\beta\cE(s)} p(s) = (1+3e^{-\eps})/4$.
Note that $\zeta^{-1}$ is in the range between $1$ and $4$, and can 
be regarded as the effective size of the nucleotide ``alphabet'' as 
``seen'' by the TF in the specific binding mode. 
The maximum value $\zeta^{-1}=4$ is attained if the energy matrix has 
infinite discrimination, $\eps\to\infty$, while no discrimination can 
be achieved at $\eps=0$ where $\zeta^{-1}=1$.
In Fig.~\ref{cond.Zsp}(a), 
we indicate the allowed region $\Zsp(\eps, L) \le 1$ in the
parameter space of $(\eps, L)$, with the boundary 
$L^*(\eps) = \ln N/\ln \zeta^{-1}(\eps)$ defined by 
$\Zsp(L^*,\eps)=1$. From the figure, it is clear that the desire
for small $\eps$ pushes the system to the boundary at $\Zsp=1$.
Along the boundary, the smallest $\eps$ is given by the largest
allowable binding length $L$. For typical bacterial
TF's whose binding sequences are no longer than
about $15$ bp (usually dimers), we find $\eps \approx 2$.

While the result on $\eps$ is somewhat specific to the two-state model, 
the need for $\Zsp \to 1$ imposed by the programmability 
consideration forces the threshold energy to take on the value
\begin{equation}
\Eth = \betainv \ln N \approx 16\,\betainv
\label{cond.Eth}
\end{equation}
(for $N\sim 10^7$) according to the condition (\ref{cond.1TF}) 
independent of the specifics of the binding energy matrix $\cE$. 
It also follows that
\begin{equation}
\Fb \approx 0
\label{cond.Fb}
\end{equation}
so that the binding threshold is simply given by 
\begin{equation}
\nt[\vS_t] \approx e^{\beta E[\vS_t]}\;.
\label{nt-set}
\end{equation}
The dependences of the $\nt$ on the number of mismatches for the 
two-state model are shown in Fig.~\ref{cond.Zsp}(b). 
We see that at the optimal parameter choice of $(\eps=2, L=15)$, each 
mismatch increases
the binding threshold $\nt$ by nearly 10-fold. In principle, further
fine-tuning can be accomplished by utilizing small variations in 
the mismatch energies.

\section*{Discussion}
\begin{figure}
\includegraphics[width=5.5cm]{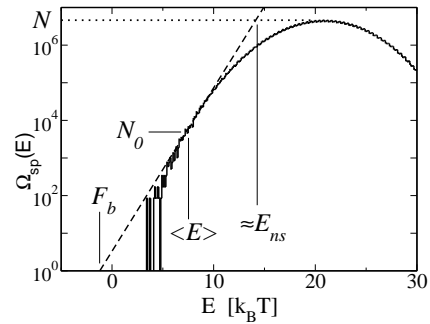}
\caption{\label{dos}
Graphical construction of the background free energy $\Fb$ and other 
quantities used in the text. 
}
\end{figure}

The key results of this study, that maximal programmability of the 
binding threshold $\nt$ requires the TF-DNA interaction 
to satisfy the conditions (\ref{cond.Eth}) and (\ref{cond.Fb}),
can be conveniently summarized graphically using the density
of states $\dos(E)$. In Fig.~\ref{dos}, the density of states is plotted
with the normalization that $\max_E\dos(E) = N$, as indicated by
the horizontal dotted line. The background
free energy $\Fb$ can be obtained 
using the Legendre construction: One draws the line $e^{\beta(E-\Fb)}$
(the dashed line in the semi-log plot of Fig.~\ref{dos}) such 
that it just
touches $\dos(E)$. $\Fb$ can then be read off as the intercept of the
dashed line on the $E$-axis, which should be in the vicinity of the origin
according to (\ref{cond.Fb}). Similarly, $\Eth$ [as given by
(\ref{cond.Eth})] can be read off 
as the $E$-coordinate where the dashed line intersects the horizontal
dotted line.

The point where the dashed line tangents $\dos(E)$ is also physically 
meaningful: The $E$-coordinate
of the tangent point gives the ensemble-averaged binding energy
$E_0\equiv\sum_E\, E\,\dos(E)\, e^{-\beta E}/Z_\mathrm{sp}$.
The vertical coordinate $N_0$ of the tangent point is given by
the relation $\Fb =  E_0 - \betainv \ln N_0$, which expresses the fact 
that the dominant contribution to the background free energy stems from 
the $N_0$ sequences of energy $\approx E_0$ in the collection of $N$ 
random sequences: The Boltzmann
weight of those sequences with $E>E_0$ is too small to contribute 
to the partition sum, while for $E<E_0$, there are too few sequences.

The value of $N_0$ is an important characteristics of the system.
$S=\ln N_0$ is known as the ``entropy'' of this system,
and $H=\ln(N/N_0)$ is known as the ``relative entropy''; the latter has been 
used to characterize the specificity of the TF-DNA interaction 
\cite{stormo}. 
As mentioned before, the annealed approximation is only valid if many 
terms contribute to the partition sum, i.e., if $N_0\gg 1$ or $S>0$. 
For the two-state model (\ref{twostate}), the values of $\eps$ and $L$ 
corresponding to the line $N_0=1$ is far from the line $L^*(\eps)$ selected
by the maximal programmability criterion; this justifies the use
of the annealed approximation. 
At the optimal parameter of $\eps=2$ and $L=15$, we have 
$N_0 \approx 10^3 \gg 1$.
The corresponding relative entropy is $H \approx 7$ (about $10$ bits).

The large value of $N_0$ also provides us with an intuitive understanding
of the simple dependence (\ref{mu.1}) of the chemical potential $\mu$ on 
the cellular TF concentration $n$; see Fig.~\ref{fig2}(b). 
As mentioned already, the expression (\ref{mu.1}) is obtained if multiple 
occupancy of the background sequences is negligible at the TF 
concentration $n$. 
Since there is a large number (i.e., $N_0$) binding sequences which 
contribute significantly to the net effect of background binding, multiple 
occupancy of these sequences is indeed not likely if $n < N_0$. 
Thus for $N_0 \sim O(10^3)$, the expression (\ref{mu.1}) can be
taken as a good approximation of the chemical potential over the 
typical range of cellular TF concentration $n=1\ldots 10^3$,
as shown in Fig.~\ref{fig2}(b) for the 3 known TF's.
We expect this result to hold even if there are multiple target
sequences, say $m_t$, whose binding energy $E_t$ is much lower than 
$E_0$, as long as $E_t > \betainv\,\ln m_t$ such that $\Fb$ is not 
affected by the addition of these target sequences to the density of states.
Having $\mu(n)$ independent of the number of targets is a 
desirable functional robustness property from a system perspective, since
one wouldn't want to perturb the recognition of the TF's and the
existing targets by the addition of a few new targets. It will be
interesting to see to what extent this feature is preserved by studying
the energetics of TF's with a large number of target sites, e.g., the
catabolic repressor protein CRP in \ecoli\ \cite{neidhardt}. 

Finally, we compare the values of the optimal interaction parameters
according to our theory to those of the well-studied TF's. From the values
listed in Table~\ref{compare}, we see that all of the available data is 
in the neighborhood of the expectation based on the maximal programmability
criterion. 
We do not suggest here that programmability was necessarily the selective 
driving force that constrained the TF-DNA interaction to its observed 
form (there could be other reasons, e.g. biochemical restrictions, for the 
interaction to be of this form). 
However, the rough correspondence between theory and observation does
indicate that it is {\em possible} (and perhaps even very likely) that
TF's generally have the required energetics for 
their binding threshold to be programmable over a wide range.

\begin{table}
\begin{tabular}{l|c|c|c|c|c}
 & \hspace*{0.2cm} {\bf theory}\hspace*{0.2cm} 
 & \hspace*{0.2cm} {\it Mnt}\hspace*{0.2cm} &\hspace*{0.2cm} 
{\it Cro}\hspace*{0.2cm} 
&\hspace*{0.2cm} {\it cI} \hspace*{0.2cm} & \hspace*{0.2cm}
{\it LacR}\hspace*{0.2cm} \\
\hline
$\Fb\; [\betainv]$ & {\bf 0} & -1.2 & -1.6 & -0.8 & -- \\
\hline
$H$ [bits] & $\approx\,${\bf 10} & 8.9 & 13.5 & 12.7 & -- \\
\hline
$\Eth\; [\betainv]$ & {\bf 16} & 17$^\dagger$ & -- & -- & $\approx 16$
\end{tabular}
\caption{\protect\flushing 
Comparison of the expected values of the background free energy $\Fb$,
relative entropy $H$ and the threshold to nonspecific binding $\Eth$ 
to the known values of these parameters for
{\it Mnt}, {\it Cro}, the $\lambda$-repressor {\it cI}, 
and the {\it lac}-repressor {\it LacR}.
The units of these values are given by the bracket in the first column.
A dash indicates that the value is not available.\\
$^\dagger$see Ref.~\protect\cite{sauer}}
\label{compare}
\end{table}

One obvious short-coming of the above comparison is that the three 
TF's for which the interaction parameters are known
are all from bacteriophages and may not represent typical prokaryotic 
TF's. 
It will therefore be very important to experimentally determine the
interaction parameters for a variety of different TF's. 
The results of a sufficient number of such studies will inform us 
whether programmability is a generic feature of TF-DNA interaction. 
Knowledge of this kind can be very helpful in developing
appropriate coarse-grained models of gene regulation at the system level.
In particular, quantitative relations of the type suggested by
Eq.~(\ref{nt-set}) will be necessary for an eventual quantitative 
description of gene regulatory networks. 
Also, this knowledge would have important implications for the 
evolution of gene regulation \cite{shraiman,us}. 

{\bf Acknowledgments. ---} 
It is a pleasure to acknowledge useful discussions with 
G.~Stormo, P.~von~Hippel, and K.~Sneppen  on many aspects of 
TF-DNA interaction. We are also grateful to the hospitality of
the Institute for Theoretical Physics in Santa Barbara where 
some of the work was carried out.
This research is supported in part by the National Science Foundation 
through Grant Nos. DMR-9971456. UG was supported in part by
a fellowship from the `Deutscher Akademischer Austauschdienst', and 
TH by a Burroghs-Wellcome functional genomics award.

\begin{thebibliography}{99}
%
\bibitem{davidson02}
Davidson, E.H. {\it et al.} (2002) 
{\it Science} {\bf 295}, 1669--1677.
%
\bibitem{eisen02}
Berman, B.P. {\it et al.} (2002) 
{\it Proc. Natl. Acad. Sci. USA} {\bf 99}, 757--762.
%
\bibitem{siggia01}
Bussemaker, H.J., Li, H. \& Siggia, E.D. (2001) 
{\it Nature Genet.} {\bf 27}, 167--171.
%
\bibitem{davidson01}
Yuh, C., Bolouri, H. \& Davidson, E.H. (2001) 
{\it Development} {\bf 128}, 617--629.
%
\bibitem{neidhardt}
Neidhardt, F.C., ed. (1996)
{\it Escherichia coli and Salmonella: cellular and molecular biology}
(ASM Press, Washington D.C.)
%
\bibitem{ptashne}
Ptashne, M. \& Gann, A. (2002) 
{\it Genes \& Signals} (Cold Spring Harbor Laboratory Press, 
Cold Spring Harbor, N.Y.)
%
\bibitem{alon}
Kalir, S., McClure, J., Pabbaraju, K., Southward, C., Ronen, M., 
Leibler, S., Surette, M.G. \& Alon, U. (2001) 
{\it Science} {\bf 292}, 2080--2083.
%
\bibitem{berg81}
Berg, O.G., Winter, R.B. \& von Hippel, P.H. (1981) 
{\it Biochemistry} {\bf 20}, 6929--6948.
%
\bibitem{winter81a}
Winter, R.B. \& von Hippel, P.H. (1981) 
{\it Biochemistry} {\bf 20}, 6948--6960.
%
\bibitem{winter81b}
Winter, R.B., Berg, O.G. \& von Hippel, P.H. (1981) 
{\it Biochemistry} {\bf 20}, 6961--6977.
%
\bibitem{vonhippel86}
von Hippel, P.H. \& Berg, O.G. (1986) 
{\it Proc. Natl. Acad. Sci. USA} {\bf 83}, 1608--1612.
%
\bibitem{berg87}
Berg, O.G. \& von Hippel, P.H. (1987) 
{\it J. Mol. Biol.} {\bf 193}, 723--750. 
%
\bibitem{dehaseth77}
deHaseth, P.L., Gross, C.A., Burgess, R.R. \& Record, M.T. Jr. (1977) 
{\it Biochemistry} {\bf 16}, 4777-4783.
%
\bibitem{record77}
Record, M.T., Jr., deHaseth, P.L. \& Lohman, T.M. (1977) 
{\it Biochemistry} {\bf 16}, 4791--4796. 
%
\bibitem{kaohuang77}
Kao-Huang, Y., Revzin, A., Butler, A.P., O'Conner, P., Noble, D.W. 
\& von Hippel, P.H. (1977) 
{\it Proc. Natl. Acad. Sci. USA} {\bf 74}, 4228--4232. 
%
\bibitem{fields}
Fields, D.S., He, Y., Al-Uzri, A.Y. \& Stormo, G.D. (1997) 
{\it J. Mol. Biol.} {\bf 271}, 178--194.
%
\bibitem{stormo}
Stormo, G.D. \& Fields, D.S. (1998) 
{\it Trends in Biochemical Sciences} {\bf 23}, 109--113.
%
\bibitem{sarai}
Sarai, A. \& Takeda, Y. (1989) 
{\it Proc. Natl. Acad. Sci. USA} {\bf 86}, 6513--6517.
%
\bibitem{takeda}
Takeda, Y., Sarai, A. \& Rivera, V.M. (1989) 
{\it Proc. Natl. Acad. Sci. USA} {\bf 86}, 439--443.
%
\bibitem{frank97}
Frank, D.E., Saecker, R.M., Bond, J.P., Capp, M.W., Tsodikov, O.V., 
Melcher, S.E., Levandoski, M.M. \& Record, M.T., Jr. (1997) 
{\it J. Mol. Biol.} {\bf 267}, 1186--1206. 
%
\bibitem{pabo92}
Pabo, C.O. \& Sauer, R.T. (1992) 
{\it Annu. Rev. Biochem.} {\bf 61}, 1053--1095. 
%
\bibitem{derrida81}
Derrida, B. (1981) {\it Phys. Rev. B} {\bf 24}, 2613--2626.
%
\bibitem{bryngelson}
Bryngelson, J.D. \& Wolynes, P.G. (1987) 
{\it Proc. Natl. Acad. Sci. USA} {\bf 84}, 7524--7528. 
%
\bibitem{droge}
Droge, P. \& Muller-Hill, B. (2001) 
{\it Bioessays} {\bf 23}, 179--183. 
%
\bibitem{elowitz99}
Elowitz, M.B., Surette, G.S., Wolf, P.-E., Stock, J.B. \& Leibler, S. (1999) 
{\it J. Bacteriol.} {\bf 181}, 197--203. 
%
\bibitem{shraiman}
Sengupta, A.M., Djordjevic, M. \& Shraiman, B.I. (2002) 
{\it Proc. Natl. Acad. Sci. USA} {\bf 99}, 2072--2077.
%
\bibitem{us}
Gerland, U. \& Hwa, T. (2002) 
{\it J. Mol. Evol.} {\bf 55}, 386--400.
%
\bibitem{sauer}
Raumann, B.E., Knight, K.L. \& Sauer, R.T. (1995) 
{\it Nat. Struct. Biol.} {\bf 2}, 1115--1122. 
%
\end{thebibliography}
\end{document}